\documentclass[journal]{IEEEtran}
\usepackage{graphicx}
\usepackage{amsmath}
\usepackage{amssymb}

\newtheorem{theorem}{Theorem}

\newtheorem{proposition}{Proposition}
\newtheorem{definition}{Definition}
\newtheorem{corollary}{Corollary}
\newlength{\shortcomment}
\newlength{\examplewidth}

\begin{document}

\title{Thresholds of Spatially Coupled Systems via Lyapunov's Method}
\author{Christian~Schlegel and Marat Burnashev%
%\thanks{Manuscript submitted January, 2004, revised January 2005, May 2005, September 2005,
%March 2006, and April 2006.}%
\thanks{Christian Schlegel is with the
Department of Electrical and Computer Engineering, Dalhousie
University, Halifax, Canada.}
\thanks{Marat Burnashev is with the Institute for Information
Transmission Problems, Russian Academy of Sciences, Moscow, Russia}
\thanks{Supported in part by NSERC Canada and by the Russian Fund
for Fundamental Research (project number 12-01-00905a).}}
\maketitle

\begin{abstract}
The threshold, or saturation phenomenon of spatially coupled systems is revisited
in the light of Lyapunov's theory of dynamical systems. It is shown that an application
of Lyapunov's direct method can be used to quantitatively describe the threshold
phenomenon, prove convergence, and compute threshold values. This provides a general
proof methodology for the various systems recently studied. Examples of spatially coupled
systems are given and their thresholds are computed.
\end{abstract}


\section{Introduction}

In this paper we apply Lyapunov's classic theory \cite{L1} to the
case of spatially coupled information processing systems, and show
that the recently proposed ``potential functions'' used in the proofs
in \cite{TTK1,YJnp1,YJnp2} are, in fact, an example from a wide class
of Lyapunov functions. Such a systematic approach to the
problem provides a general tool to deal with the dynamics of
spatial coupling.
The required definitions and the Lyapunov theorem are described below.

Iterative signal and information processing has enjoyed a tremendous
rise in popularity with the introduction of turbo coding
\cite{BerGla96}, and various ``statistical'' analysis methods
have been developed to study the performance of iterative processors,
in particular the method of extrinsic information exchange (EXIT)
introduced by ten Brink \cite{tenB003}, variance transformation
by Divsalar et.~al.~\cite{DivDolPol01}, and density evolution (DE),
refined by Richardson and Urbanke \cite{RicUrb01}.

Spatial coupling emerged in the information processing arena largely
by ``accident'', and in the form of low-density parity-check (LDPC)
convolutional coding \cite{FelZig1999}. Researchers noted that these
codes could be designed with decoding thresholds that are very close
to the channel capacity. The effect of spatial coupling derives from
the special structure of these codes, where a large set of random
codes are linked in a controlled fashion. The performance advantage
comes from ``anchoring'' initial symbols to known values on one (or both) 
side(s) of this chain of linked codes, which causes a locally
smaller rate and accelerated convergence. This in turn allows the
entire code to converge at signal-to-noise ratios where uniform
convergence is otherwise not possible. Recently, it has been shown
that spatial coupling can decrease the convergence threshold in
low-density parity-check codes on binary-erasure channels all the way
to the maximum-likelihood decoding threshold \cite{SpaCoup2010}, a
phenomenon knows as ''threshold saturation''.

This phenomenon has given rise to much research activity in attempting
to use this effect to show optimal performance for certain coupled
communications and coding systems \cite{SchTru2012}, and to find
general proof methodologies for analyzing spatially coupled systems
\cite{RU1,TTK1,YJnp1,YJnp2}.

\section{The System}
\subsection{Basic Dynamical Systems}

We consider a {\it discrete dynamical system},
governed by the following iteration equation
\begin{equation}\label{basic1}
\boldsymbol{x}^{(l+1)} = \boldsymbol{f}\left( \boldsymbol{g}
\left(\boldsymbol{x}^{(l)} \right);
\varepsilon\right), \qquad  l=0,1,2,\ldots,
\end{equation}
where $\boldsymbol{x} \in \mathcal{X} = [0,1]^{d} \subset
\mathbb{R}^{d}$, $\varepsilon \in {\mathcal E} = [0,1]$ and
$\boldsymbol{f}: \mathcal{X} \times \mathcal{E} \rightarrow
\mathcal{X}$ is a sufficiently smooth function. Also assume that
$\boldsymbol{f}\left(\boldsymbol{z};\varepsilon\right)$
and $\boldsymbol{g}\left(\boldsymbol{x};\varepsilon\right)$
are strictly increasing in both arguments, and that also
$\boldsymbol{f}(\mathbf{0};\varepsilon) = \boldsymbol{f}
(\boldsymbol{g} \left(\boldsymbol{x}\right) ;0) = \mathbf{0}$.

The {\em single-system} dynamical equation corresponds to the situation
where
\begin{equation}
x_i^{(l+1)} = f \left( g \left( x_i^{(l)} \right); \varepsilon \right), \hspace{5mm}
i=1,\cdots, d.
\label{basic2}
\end{equation}
The system (\ref{basic2}) can represent the convergence properties of
an LDPC decoder, for example \cite{SchPer04}, the variance evolution of
an iterative cancelation receiver \cite{SchTru2012}, or similar systems described
by bi-partite (Tanner) graphs, and the functions $f(\cdot)$ and $g(\cdot)$ describe the statistical
behavior of key performance parameters of the two types of processing nodes in these graphs. 
Such equations are typically obtained by applying a density evolution analysis to
the system in question. In the case where $\boldsymbol{f}(\cdot)$ and $\boldsymbol{g}(\cdot)$
are vector functions (\ref{basic1}) describes the evolution of {\em spatially coupled} systems,
where the joint behavior of all $d$ sub-systems needs to be studied.

In this context, one is typically interested in the largest $\varepsilon$ such that
for any $\boldsymbol{x} \in \mathcal{X}$
$\lim\limits_{l \to \infty} \boldsymbol{x}^{(l)} = {\mathbf 0}$. This parameter
is typically a signal-to-noise ratio \cite{SchTru2012}, or a channel error rate
in the case of LDPC codes \cite{RU1}.

With $\varepsilon \in {\mathcal E}$, and
$\boldsymbol{x}^{(0)} = \boldsymbol{x} \in \mathcal{X}$, let
\begin{equation}\label{defx1}
\boldsymbol{x}^{\infty}(\boldsymbol{x};\varepsilon) =
\lim\limits_{l \to \infty}\boldsymbol{x}^{(l)} =
\lim\limits_{l \to \infty}\boldsymbol{f}\left( \boldsymbol{g}\left( \boldsymbol{x}^{(l)} \right);
\varepsilon\right).
\end{equation}
This limit exists for
all $\varepsilon \in {\mathcal E}$ due to the monotonicity
of $\boldsymbol{f}$, and therefore of $\boldsymbol{x}^{(l)}$ in $l$
(see \cite[Lemma 15]{SpaCoup2010}, \cite[Lemma~2]{YJnp2}).
\medskip

We will need the following
\begin{definition}
The {\sl single-system \eqref{basic2} threshold}
is defined as
\begin{equation}\label{sst}
\varepsilon^{*}_{\mathbf s}= \sup\left\{\varepsilon \! \in \! {\mathcal E}|
\boldsymbol{x}^{\infty}({\mathbf 1};\varepsilon) \! = \! {\mathbf 0}
\right\}
= \sup\left\{\varepsilon\! \in\! {\mathcal E}| x^{\infty}(1;\varepsilon) \! = \! {0}
\right\}.
\end{equation}
\end{definition}
\medskip

The threshold $\varepsilon^{*}_{\mathbf s}$ is the well-known threshold
of iterative decoders and demodulators as discussed amply in the literature.
It can be computed by elementary methods applied to the single-variable
dynamical system \eqref{basic2}. 

In the sequel of this paper we will focus on coupled systems of the type \eqref{basic1}.

\subsection{Coupled Dynamical Systems}

We start with a basic (``$1$-dimensional'') system \eqref{basic1} with the state-space
$\mathcal{X}$. Assume that we have $L$ identical independent copies
of this $1$-dimensional system. Together they form an $L$-dimensional
system \eqref{basic1} with state-space $\mathcal{X}^{L}$. If nothing
else is done, the fixed points of that $L$-dimensional system coincide with
fixed points of the original $1$-dimensional system.

Now, without enlarging the space $\mathcal{X}^{L}$ imagine that these $L$ identical
systems are arranged in a linear fashion from left to right, for example, and therefore there are
two boundaries. We now introduce dependencies for each of the $L$ systems on its
$w$ adjacent neighboring systems. These dependencies shall be identical, when possible,
for all $L$ systems. The only exception will be for systems close to a boundary.
If some connection is not possible, it is assumed to be connected to a known value (this is
the anchor value). As a result the overall system now possesses a boundary
asymmetry, which will imply additional properties. As experience with spatial coupling
has shown, this asymmetry, in the form of the known values starting at the boundary systems,
slowly propagates with iterations to the inner systems. If coupling is strong enough then
iterations can remove all non-zero fixed points of the overall system and achieve improved
convergence thresholds over the $1$-dimensional system.

Specifically, the coupled discrete dynamical system is then described by the
iteration equations \cite{SpaCoup2010}
\begin{equation}\label{22coup1}
x_{i}^{(l+1)} = f\left(g\left(\frac{1}{w^{2}}\sum\limits_{k=0}^{w-1}
\sum\limits_{j=0}^{w-1}
x_{i+j-k}^{(l)}\right);\varepsilon\right).
\end{equation}

{\sl Remark 1}. In \cite{YJnp1, YJnp2}  the following system similar to
\eqref{22coup1} was considered:
\begin{equation}\label{2coup1}
x_{i}^{(l+1)} = \frac{1}{w}\sum\limits_{k=0}^{w-1}
f\left(\frac{1}{w}\sum\limits_{j=0}^{w-1}
g\left(x_{i+j-k}^{(l)}\right);\varepsilon\right).
\end{equation}
Both systems \eqref{22coup1} and \eqref{2coup1} have the same
threshold and can be analyzed by similar methods. We consider
the system \eqref{22coup1} in the sequel.
Note that if the function $f(x)$ is $\cup$-convex (as is usually the case)
then
$$
\begin{gathered}
x_{i}^{(l+1)} = f\left(g\left(\frac{1}{w^{2}}\sum\limits_{k=0}^{w-1}
\sum\limits_{j=0}^{w-1}
x_{i+j-k}^{(l)}\right);\varepsilon\right) \leq \\
\leq \frac{1}{w}\sum\limits_{k=0}^{w-1}
f\left(\frac{1}{w}\sum\limits_{j=0}^{w-1}
g\left(x_{i+j-k}^{(l)}\right);\varepsilon\right),
\end{gathered}
$$
and therefore the system \eqref{22coup1} has convergence
properties that are no worse then those of \eqref{2coup1}.

A way to investigate the system \eqref{basic1} was
offered in \cite{TTK1} and developed in \cite{YJnp1, YJnp2}.
It is based on using the following function
$U(\boldsymbol{x}): \boldsymbol{x} \rightarrow \mathbb{R}^{1}$,
called the {\sl potential function}\footnote{This terminology stems
from the fact that the integral in \eqref{defpf} does not depend on the curve ${\cal C}$ from
${\mathbf 0}$ to $\boldsymbol{x}$ along which this integral is computed.},
\begin{equation}\label{defpf}
U(\boldsymbol{x}) = \int\limits_{{\mathbf 0}}^{\boldsymbol{x}}
\boldsymbol{g}'(\boldsymbol{z})\left[\boldsymbol{z} -
\boldsymbol{f}\left(\boldsymbol{g}(
\boldsymbol{z})\right)\right]d\boldsymbol{z}
\end{equation}
When the vector system
considered is constructed from one-dimensional systems as in
\eqref{22coup1} or \eqref{2coup1}, definition \eqref{defpf}
reduces to the one-dimensional function
$U(x): x \rightarrow \mathbb{R}^{1}$:
\begin{equation}\label{defpf11}
U(x) = \int\limits_{0}^{x}g'(z)\left[z - f(g(z))\right]dz.
\end{equation}

The motivation for using the function $U(\boldsymbol{x})$  in \cite{TTK1} 
was based on a continuous-time approximation for the system
\eqref{2coup1}, given by
\begin{equation}\label{contapp1}
\frac{d\boldsymbol{x}(t)}{dt} =
\boldsymbol{f}\left(\boldsymbol{g}(\boldsymbol{x}(t))\right) -
\boldsymbol{x}(t), \qquad  t > 0, \quad t \to \infty,
\end{equation}
and, in turn, on the close relation of an analog of the function
$U(\boldsymbol{x})$ for the system \eqref{contapp1} to its Bethe
free energy.

The main aim of the paper is to give another (more traditional) look
at the problem considered based on using {\sl Lyapunov functions}.
We show that from that point of view the function $U(\boldsymbol{x})$
from \eqref{defpf} is, in fact, an example from a wide class of
{\sl Lyapunov functions} for the system \eqref{basic1}, constructed
by the {\sl variable gradient} method \cite[Chapter 3.4]{HC}.

\section{Lyapunov Formulation}

\subsection{Lyapunov's Direct Method}

Essentially, Lyapunov built a theory whereby the often exceedingly complicated
study of when and how dynamical systems converge is moved away from studying
the behavior of individual trajectories to studying the behavior of the system in
certain regions of space. This is simplified by studying {\em Lyapunov candidate functions}
in these regions.
\medskip

\begin{definition}
The solution $\boldsymbol{x}^{(l)} \equiv {\mathbf 0}$ to
\eqref{basic1} is {\sl globally asymptotically stable} if
$\lim\limits_{l \to \infty} \boldsymbol{x}^{(l)} = {\mathbf 0}$
for all $\boldsymbol{x}^{(0)} \in {\mathcal X}$.
\end{definition}
\medskip

Denote by ${\mathcal X}_{0} = {\mathcal X} \setminus \{{\mathbf 0}$\};
${\mathcal E}_{0} = {\mathcal E} \setminus \{0\}$, and let
${\mathcal E}_{2} \subseteq {\mathcal E}_{0}$ be a subset of
${\mathcal E}_{0}$.

\medskip
A Lyapunov candidate function is defined in
\begin{definition}
A continuous function
$V(\boldsymbol{x};\varepsilon): \mathcal{X} \times \mathcal{E}_{2}
\rightarrow \mathbb{R}^{1}$ is called a {\sl Lyapunov function}
for the system \eqref{basic1} with
$\varepsilon \in {\mathcal E}_{2}$, if it satisfies the following
conditions:
\begin{eqnarray}\label{Lcond1}
V({\mathbf 0};\varepsilon) & = & 0, \quad \varepsilon \in {\mathcal E}_{2}; \\
\label{Lcond2}
V(\boldsymbol{x};\varepsilon) & > &  0, \quad \boldsymbol{x} \in
{\mathcal X}_{0}, \varepsilon \in {\mathcal E}_{2}; \\
\label{Lcond3}
V(\boldsymbol{f}(\boldsymbol{x};\varepsilon);\varepsilon) -
V(\boldsymbol{x};\varepsilon) & < & 0, \quad \boldsymbol{x} \in
{\mathcal X}, \varepsilon \in {\mathcal E}_{2}.
\end{eqnarray}
\end{definition}
\medskip

The following result, known as Lyapunov's direct method (1892),
gives sufficient conditions for global asymptotic stability of
the system \eqref{basic1}.
\medskip

\begin{theorem}\label{th:one}
(modification of \cite[Theorem 13.2]{HC}).
Assume that $V(\boldsymbol{x};\varepsilon)$ is a Lyapunov function
for the system \eqref{basic1} with $\varepsilon \in {\mathcal E}_{2}$.
Then the solution $\boldsymbol{x}^{(l)} \equiv {\mathbf 0}$ is
globally asymptotically stable.
\end{theorem}
\medskip

\subsection{Lypunov Function of Spatially Coupled Systems}

Represent the system \eqref{basic1} in the form
\begin{equation}\label{basic42a1}
\boldsymbol{z}^{(l)} - \boldsymbol{z}^{(l+1)} =
\boldsymbol{q}\left(\boldsymbol{z}^{(l)}\right),
\qquad  l=0,1,2,\ldots.
\end{equation}
where $\boldsymbol{q}(\boldsymbol{z}) = \boldsymbol{z} -
\boldsymbol{f}(\boldsymbol{g}(\boldsymbol{z}))$.
In order to have convergence
$\boldsymbol{z}^{(l)} \to \boldsymbol{z}^{\infty}$ it is sufficient
that the condition 
\begin{equation}\label{basic42b1}
\boldsymbol{q}\left(\boldsymbol{z}^{(l)}\right) \geq \mathbf{0},
\quad l=0,1,2,\ldots
\end{equation}
is fulfilled along the trajectory $\boldsymbol{z}^{(0)},\boldsymbol{z}^{(1)},
\boldsymbol{z}^{(2)}, \ldots$.
In order to have convergence $\boldsymbol{z}^{(l)} \to \mathbf{0}$ it
is sufficient, in addition to \eqref{basic42b1}, that the following
condition is fulfilled:

For any $\delta > 0$ there exists $\varepsilon = \varepsilon(\delta) > 0$
such that
\begin{equation}\label{basic42b11}
\left\|\boldsymbol{q}\left(\boldsymbol{z}^{(l)}\right)\right\| \geq
\varepsilon(\delta), \quad \mbox{\rm if } \quad
\left\|\boldsymbol{z}^{(l)}\right\| \geq \delta.
\end{equation}

We want to avoid dealing with the trajectory
$\{\boldsymbol{z}^{(l)}\}$ and replace condition
\eqref{basic42b11} by a simpler check. For
that purpose the following auxiliary result is useful. The system
\eqref{basic42a1} is similar to the continuous-time system
\begin{equation}\label{basic42c1}
\frac{d\boldsymbol{z}(t)}{dt} = -\boldsymbol{q}(\boldsymbol{z}(t)),
\qquad  t > 0, \quad t \to \infty,
\end{equation}
where $\boldsymbol{q}(\boldsymbol{z}(t)) = \boldsymbol{z}(t) -
\boldsymbol{f}(\boldsymbol{g}\left(\boldsymbol{z}(t)\right))$.

To approach this problem systematically we apply  the {\sl variable gradient} 
method for constructing Lyapunov functions to the system \eqref{basic42c1}, 
\cite[Chapter 3.4]{HC}. It will be a Lyapunov function for the system \eqref{basic1} 
as well.

Let $V: \mathcal{Z} \rightarrow \mathbb{R}^{1}$ be a
continuously differentiable function and let
$$
\boldsymbol{h}(\boldsymbol{z}) =
\left(\frac{\partial V}{\partial \boldsymbol{z}}\right)^{\top},
$$
i.e. $\boldsymbol{h}(\boldsymbol{z})$ is the gradient of
$V(\boldsymbol{z})$. Here
$$
\begin{gathered}
\frac{\partial V}{\partial \boldsymbol{z}} = \left[
\frac{\partial V}{\partial z_{1}}, \frac{\partial V}{\partial z_{2}},
\ldots, \frac{\partial V}{\partial z_{n}}\right] - \mbox{row-vector},
\\ \boldsymbol{h}(\boldsymbol{z}) - \mbox{column-vector}.
\end{gathered}
$$
The derivative of $V(\boldsymbol{z})$ along the
trajectories of \eqref{basic42c1} is given by
\begin{equation}\label{basic42d}
\frac{dV(\boldsymbol{z})}{dt} = -\frac{\partial V}
{\partial \boldsymbol{z}}\boldsymbol{q}(\boldsymbol{z}) =
-\boldsymbol{h}^{\top}(\boldsymbol{z})\boldsymbol{q}(\boldsymbol{z}).
\end{equation}

Next, construct $\boldsymbol{h}(\boldsymbol{z})$ such that
$\boldsymbol{h}(\boldsymbol{z})$ is a gradient for a positive
function and
\begin{equation}\label{basic42d1}
\frac{dV(\boldsymbol{z})}{dt} = -\boldsymbol{h}^{\top}(\boldsymbol{z})
\boldsymbol{q}(\boldsymbol{z}) < 0, \qquad
\boldsymbol{z} \in \mathcal{Z}, \quad \boldsymbol{z} \neq \mathbf{0}.
\end{equation}
Specifically, the function $V(\boldsymbol{z})$ can be computed from
the line integral
\begin{equation}\label{basic42e}
V(\boldsymbol{z}) = \int\limits_{\mathbf{0}}^{\boldsymbol{z}}
\boldsymbol{h}^{\top}(\boldsymbol{s})d\boldsymbol{s}.
\end{equation}
Recall that the line integral of a gradient vector
$\boldsymbol{h}: \mathbb{R}^{n} \rightarrow \mathbb{R}^{n}$ is path
independent, and hence, integration in \eqref{basic42e} can
be taken along any path joining the origin to
$\boldsymbol{z} \in \mathcal{Z}$.

It is known \cite[Proposition 3.1]{HC} that
$\boldsymbol{h}(\boldsymbol{z})$ is a gradient of a real-valued
function $V: \mathbb{R}^{n} \rightarrow \mathbb{R}^{1}$ if and only
if the Jacobian matrix
$\partial \boldsymbol{h}/\partial \boldsymbol{z}$ is
symmetric, i.e. iff
\begin{equation}\label{basic42f}
\frac{\partial h_{i}}{\partial z_{j}} =
\frac{\partial h_{j}}{\partial z_{i}}, \qquad i,j = 1,\ldots,n.
\end{equation}

According to the definition of Lyapunov function,
choosing $\boldsymbol{h}(\boldsymbol{z})$ and arriving at
$V(\boldsymbol{z})$, it is necessary to have
\begin{equation}\label{basic42g1}
V(\boldsymbol{z}) > 0, \qquad \boldsymbol{z} \in \mathcal{Z}_{0},\;
\boldsymbol{z} \neq \mathbf{0},
\end{equation}
where $\mathcal{Z}_{0} \subseteq \mathcal{Z}$ is any open set such
that $\boldsymbol{z}(t) \in \mathcal{Z}_{0}$ for all $t > 0$. The
larger the set $\mathcal{Z}_{0}$ we can find (based, perhaps, on
additional information about $\boldsymbol{z}(t)$), the less restrictive
is condition \eqref{basic42g1}.

Referring to \eqref{basic42d1} we look for
$\boldsymbol{h}(\boldsymbol{z})$ of the form
$\boldsymbol{h}(\boldsymbol{z}) = {\mathbf B}(\boldsymbol{z})
\boldsymbol{q}(\boldsymbol{z})$, where ${\mathbf B}(\boldsymbol{z})$
is an $n \times n$-positive-definite matrix. Then due to
\eqref{basic42d1} we need ($\boldsymbol{z} \in \mathcal{Z}_{0}$,
$\boldsymbol{z} \neq \mathbf{0}$)
\begin{equation}\label{basic42g}
\boldsymbol{h}^{\top}(\boldsymbol{z})\boldsymbol{q}(\boldsymbol{z}) =
\boldsymbol{q}^{\top}(\boldsymbol{z})
{\mathbf B}^{\top}(\boldsymbol{z})\boldsymbol{q}(\boldsymbol{z}) > 0,
\end{equation}
and from \eqref{basic42e} we have
\begin{equation}\label{basic42h}
V_{\mathbf B}(\boldsymbol{z}) =
\int\limits_{\mathbf{0}}^{\boldsymbol{z}}
\left[{\mathbf B}(\boldsymbol{s})(\boldsymbol{z} -\boldsymbol{f}
(\boldsymbol{z}))\right]^{\top}d\boldsymbol{s}.
\end{equation}

For a chosen positive-definite $n \times n$-matrix
${\mathbf B}(\boldsymbol{z})$ the function
$V_{\mathbf B}(\boldsymbol{z})$
from \eqref{basic42h} is a Lyapunov function for the system
\eqref{basic1}, if conditions
\eqref{basic42g1}--\eqref{basic42g} are satisfied. Then
\begin{equation}\label{basic42h1}
\lim\limits_{t \to \infty}\boldsymbol{z}(t) = \mathbf{0}.
\end{equation}

We now show that the function
$V_{\mathbf B}(\boldsymbol{z})$ is a Lyapunov function 
for the system \eqref{basic1}. Represent 
\eqref{basic1} (see \eqref{basic42a1}) in the form
$$
\boldsymbol{x}^{(l+1)}-  \boldsymbol{x}^{(l)}=
\boldsymbol{q}(\boldsymbol{x}^{(l)}) = \boldsymbol{x}^{(l)} -
\boldsymbol{f}\left(\boldsymbol{g}(\boldsymbol{x}^{(l)})\right),
$$
where
$$
\begin{gathered}
\boldsymbol{q}(\boldsymbol{x}) = \boldsymbol{x} -
\boldsymbol{f}\left(\boldsymbol{g}(\boldsymbol{x})\right).
\end{gathered}
$$
Condition \eqref{basic42g} takes the form
($\boldsymbol{x} \in \mathcal{X}_{0}$,
$\boldsymbol{x} \neq \mathbf{0}$)
\begin{equation}\label{basic42j}
\left[\boldsymbol{x} -
\boldsymbol{f}\left(\boldsymbol{g}(\boldsymbol{x})\right)
\right]^{\top}{\mathbf B}^{\top}(\boldsymbol{x})\left[\boldsymbol{x} -
\boldsymbol{f}\left(\boldsymbol{g}(\boldsymbol{x})\right)\right] > 0,
\end{equation}
where $\mathcal{X}_{0} \subseteq \mathcal{X}$ is any set such
that $\boldsymbol{x}^{(l)} \in \mathcal{X}_{0}$ for all $l \geq 0$.
In turn, \eqref{basic42h} takes the form
\begin{equation}\label{basic42k}
V_{\mathbf B}(\boldsymbol{z}) =
\int\limits_{\mathbf{0}}^{\boldsymbol{x}}
\left[{\mathbf B}(\boldsymbol{s})
(\boldsymbol{s} -\boldsymbol{f}(\boldsymbol{g}
(\boldsymbol{s})))\right]^{\top}d\boldsymbol{s}.
\end{equation}
Note that if we set
${\mathbf B}(\boldsymbol{x}) = \boldsymbol{g}'(\boldsymbol{x})$, then
\begin{equation}\label{basic42k1}
V_{\mathbf B}(\boldsymbol{z}) = U(\boldsymbol{x}),
\end{equation}
where $U(\boldsymbol{x})$ is the potential function from
\eqref{defpf}.

For a chosen positive-definite $n \times n$-matrix
${\mathbf B}(\boldsymbol{x})$ the function
$V(\boldsymbol{x},{\mathbf B})$ from \eqref{basic42k} is Lyapunov
function for the system \eqref{basic1}, if condition
\eqref{basic42j} is satisfied, and, moreover,
\begin{equation}\label{basic42k2}
V_{\mathbf B}(\boldsymbol{x}) > 0, \qquad
\boldsymbol{x} \in \mathcal{X}_{0}, \quad
\boldsymbol{x} \neq \mathbf{0}.
\end{equation}
We also have $V_{\mathbf B}(\boldsymbol{0})$ = 0. If both conditions
\eqref{basic42j} and \eqref{basic42k2} are satisfied, then
\begin{equation}\label{basic42k3}
\lim\limits_{l \to \infty}\boldsymbol{x}^{(l)} = \mathbf{0}.
\end{equation}

Consider the system \eqref{2coup1} and set
${\mathbf B}(\boldsymbol{x})= D\boldsymbol{g}'(\boldsymbol{x})$,
where $D$ is a positive-definite  diagonal matrix. It was shown
in \cite[Theorem 1]{YJnp1}, \cite[Theorem 1]{YJnp2} that if
condition \eqref{basic42k2} is satisfied, then the condition
\eqref{basic42j} is also satisfied (i.e. there exists the unique
fixed point $\boldsymbol{x} = \mathbf{0}$ along the trajectory).

We give another proof of a similar result for the system
\eqref{22coup1}.

\subsection{Convergence of Spatially Coupled Systems}

Having defined the Lyapunov function of the spatially coupled system
in (\ref{basic42k}), we proceed as follows: Given a certain initial
condition, in our case $x_i = 0, i<0; i>L$, represented by the
anchoring of the spatially coupled system, we find the largest
$\varepsilon \in {\mathcal E}_{2}$, such that
(\ref{Lcond1}) -- (\ref{Lcond3}) hold.

Formally, the coupled system threshold is defined in
\begin{definition}
The {\sl coupled-system \eqref{22coup1} threshold} with
$x_i = 0, i<0; i>L$ is defined as
\begin{equation}\label{sst1}
\varepsilon^{*}_{\mathbf c}= \sup\left\{\varepsilon \in
{\mathcal E}_{2}|\boldsymbol{x}^{\infty}({\mathbf 1};\varepsilon) =
{\mathbf 0}\right\}.
\end{equation}
\end{definition}
Evidently
$\varepsilon^{*}_{\mathbf c} \geq \varepsilon^{*}_{\mathbf s}$
in general, with equality if $w=0$, or, for example, if the $L$ identical 
systems are arranged in a circle such that no boundary exists
\medskip

We will also use
\begin{definition}
For a positive-definite matrix ${\mathbf B}$ the {\sl coupled-system
\eqref{22coup1} threshold} $\varepsilon_{\mathbf c}({\mathbf B})$
is defined as
\begin{equation}\label{sst2}
\varepsilon_{\mathbf c}({\mathbf B}) =
\sup\left\{\varepsilon \in {\mathcal E}_{2}|
\min_{\boldsymbol{x} \in \mathcal{X}_{0}}
V_{\mathbf B}(\boldsymbol{x}) \geq 0\right\}.
\end{equation}
\end{definition}
\medskip

For any positive-definite matrix ${\mathbf B}$ we have
\begin{equation}\label{sst3}
\varepsilon^{*}_{\mathbf c}({\mathbf B}) \leq
\varepsilon^{*}_{\mathbf c}.
\end{equation}

Let $\boldsymbol{x}_{0} = (x_{0,-L},\ldots,x_{0,0})$ be a fixed
point of \eqref{22coup1} and
$f(g(x);\varepsilon) = \varepsilon f(g(x))$.
Then $\{x_{0,i}\}$ satisfy equations
($i \in {\mathcal L}' = \{-L,-L+1,\ldots,0\}$)
\begin{equation}\label{22coup11}
x_{0,i} = \varepsilon f\left(g\left(\frac{1}{w^{2}}
\sum\limits_{k=0}^{w-1}\sum\limits_{j=0}^{w-1}x_{0,i+j-k}\right)
\right).
\end{equation}

The following theorem represents the main result of the paper.
\medskip

\begin{theorem}
There exists a function $w_{0}(f,g)$ such that
for any positive-definite matrix ${\mathbf B}$, 
$w \geq w_{0}(f,g)$, $L \geq 2w+1$ and
$\varepsilon < \varepsilon^{*}_{\mathbf c}({\mathbf B})$
the only fixed point of the system \eqref{22coup1} is
$\boldsymbol{x}_{0} = \mathbf{0}$.
\label{th:two}
\end{theorem}
\medskip

{\bf Proof}. Our proof of Theorem \ref{th:two} is different from the proofs
of \cite[Theorem 1]{YJnp1}, \cite[Theorem 1]{YJnp2}. We have
$\boldsymbol{x}^{(l+1)} < \boldsymbol{x}^{(l)}$ and
$V_{\mathbf B}\left(\boldsymbol{x}^{(l+1)}\right) <
V_{\mathbf B}\left(\boldsymbol{x}^{(l)}\right)$ for all $l \geq 0$,
and the sequence $\{\boldsymbol{x}^{(l)}\}$ converges to a fixed point
$\boldsymbol{x}_{0}$, which is the (local) minimum  of the function
$V_{\mathbf B} \left(\boldsymbol{x}\right)$, but may never reach 
the point $\boldsymbol{x}_{0}$.
Note that if $\boldsymbol{x}_{0} \neq \mathbf{0}$ is a fixed point (i.e.
$\boldsymbol{x}_{0} - \varepsilon \boldsymbol{f}\left(\boldsymbol{g}
(\boldsymbol{x}_{0})\right) = \mathbf{0}$) then
$V'_{\mathbf B} (\boldsymbol{x}_{0} ) = \mathbf{0}$
and
$$
V''_{\mathbf B}(\boldsymbol{x}_{0}) =
{\mathbf B}(\boldsymbol{x}_{0})\left[{\mathbf I}_{n} - \varepsilon
\boldsymbol{f}'\left(\boldsymbol{g}(\boldsymbol{x}_{0})\right)\right].
$$
Then it is sufficient to prove that the matrix
${\mathbf I}_{n} - \varepsilon
\boldsymbol{f}'\left(\boldsymbol{g}(\boldsymbol{x}_{0})\right)$ has a
negative eigenvalue (i.e. it is not a positive-definite matrix) and
therefore $\boldsymbol{x}_{0}$ can not be a local minimum of the
function $V_{\mathbf B}\left( \boldsymbol{x} \right)$ (all functions
are continuous). We have
$$
\boldsymbol{f}'\left(\boldsymbol{g}(\boldsymbol{x}_{0})\right) =
\boldsymbol{f}'_{\boldsymbol{g}}\boldsymbol{g}'_{\boldsymbol{x}_{0}}
{\boldsymbol{x}_{0}}' = f'_{g}g'{\boldsymbol{x}_{0}}'.
$$

Note that if $\varepsilon$ is sufficiently small then there exists
only the zero fixed point $\boldsymbol{x}_{0} = \mathbf{0}$. As
$\varepsilon$ grows it reaches some $\varepsilon_{1} > 0$ there
appear non-zero fixed point(s) $\boldsymbol{x}_{0} \neq \mathbf{0}$.
We need to show that for $\varepsilon > \varepsilon_{1}$ the matrix
${\mathbf A} = \varepsilon \boldsymbol{f}'\left(\boldsymbol{g}
(\boldsymbol{x}_{0})\right)$ has an eigenvalue greater than $1$. The
matrix ${\mathbf A}$ is non-negative (i.e. all its elements are
non-negative). Therefore its spectral radius $\rho({\mathbf A})$
equals its maximal eigenvalue. Moreover, if ${\mathbf A}$ has a
positive eigenvector (as in our case) then the corresponding
eigenvalue is $\rho({\mathbf A})$ \cite[Chapter 8]{HJ}.

If $w = 1$ then the fixed point
$\boldsymbol{x}_{0} = (x_{0},\ldots,x_{0})$ and the matrix
${\mathbf A} = \varepsilon f'(g(x_{0})) {\mathbf I}_{n}$ is diagonal
with equal diagonal elements (all that reduces to the uncoupled case).
If $w > 1$ then the fixed point
$\boldsymbol{x}_{0} = (x_{0,1},\ldots,x_{0,n})$ consists of
nondecreasing components. The $i$-th row $A_{i}$ of ${\mathbf A}$ has
the form
\begin{equation}\label{nonnegmat2}
\begin{gathered}
A_{i} = \varepsilon a_{i}D_{i}, \qquad
a_{i} = f'_{g}(y_{i})g'(y_{i}), \\
y_{i} = \frac{1}{w^{2}}
\sum\limits_{k=0}^{w-1}\sum\limits_{j=0}^{w-1}x_{0,i+j-k},
\quad D_{i} = (D_{i,0},\ldots,D_{i,n}), \\
D_{i,j} = \frac{w-|i-j|}{w^{2}}, \quad |i-j| \leq w, \\
D_{i,j} = 0, \quad |i-j| \geq w.
\end{gathered}
\end{equation}
Diagonal elements of ${\mathbf A}$ are
$\{\varepsilon a_{i}/w,\ i=1,\ldots,n\}$. For the matrix
${\mathbf D}$ of rows $\{D_{i}\}$ and the matrix ${\mathbf A}$ of
rows $\{A_{i}\}$ we have
\medskip

{\bf Lemma 2}. For any $w \geq 1$ and $L \geq 2w+1$ the matrix
${\mathbf D}$ has the maximal eigenvalue $\rho(\mathbf D) = 1$.
The matrix ${\mathbf A}$ has the maximal eigenvalue
$\rho(\mathbf A) = \varepsilon\max\limits_{i}a_{i}$.
\medskip

Therefore, if $\varepsilon > 1/\max\limits_{i}a_{i}$ then
$\rho(\mathbf A) > 1$, and ${\mathbf I}_{n} - \varepsilon
\boldsymbol{f}'\left(\boldsymbol{g}(\boldsymbol{x}_{0})\right)$
has a negative eigenvalue. 

Remember that we still have the constraint
\eqref{basic42k2}, i.e $V_{\mathbf B}(\boldsymbol{x}) > 0$,
$\boldsymbol{x} \neq \mathbf{0}$, which sets the upper bound on
$\varepsilon$. For $\varepsilon$, satisfying both constraints,
the only fixed point of the system \eqref{22coup1} is
$\boldsymbol{x}_{0} = \mathbf{0}$.

It remains to clarify the condition
$\varepsilon > 1/\max\limits_{i}a_{i}$.
We limit ourselves here to the following result.
\medskip

\begin{proposition}\label{prop:one}
There exists a function $w_{0}(f,g)$ such that
for any $w \geq w_{0}(f,g)$, $L \geq 2w+1$ and
$\varepsilon > \varepsilon^{*}_{\mathbf s}$ the matrix
${\mathbf I}_{n} - \varepsilon
\boldsymbol{f}'\left(\boldsymbol{g}(\boldsymbol{x}_{0})\right)$
has a negative eigenvalue.
\end{proposition}
\medskip

From Proposition \ref{prop:one}, the constraint \eqref{basic42k2}, and Definition 5
Theorem \ref{th:two} follows. \qquad $\Box$
\medskip

{\sl Remark} 4. It is natural to investigate the value
$\sup_{\mathbf B}\varepsilon_{\mathbf c}({\mathbf B})$, where
supremum is taken over positive-definite matrices ${\mathbf B}$.
It will be done later.

\section{Examples:}
\subsection{LDPC Codes}

Consider the traditional example of the
$(3,6)$-regular LDPC code ensemble defined by
constant degree profile $(\lambda,\rho) = (x^{2},x^{5})$
\cite{RU1, SchPer04}. Then, for the binary erasure channel
\begin{equation}\label{ex1}
x^{(l+1)} = \varepsilon f(x^{(l)}), \qquad l = 0,1,2,\ldots
\end{equation}
where
\begin{equation}\label{deff1}
f(x) = \left[1-(1-x)^{5}\right]^{2}, \qquad 0 \leq x \leq 1.
\end{equation}

Given an initial erasure probability $x^{(0)} \in [0,1]$, we wish to find all
$\varepsilon \in [0,1]$ such that $x^{(l)} \to 0$ as $l \to \infty$.

%Introduce also the function
%\begin{equation}\label{defg1}
%g(x,\varepsilon) = x - \varepsilon f(x).
%\end{equation}
%which is a Lyapunov candidate function for this case, and it fulfills
%(\ref{Lcond1}) -- (\ref{Lcond3}) 
Now, the single system converges for all $\varepsilon < \varepsilon_0$,
where
\begin{equation}\label{1root2}
\varepsilon_{0} = \min_{0 < x < 1}\frac{x}{f(x)} = \min_{0 < x < 1}
\frac{x}{\left[1-(1-x)^{5}\right]^{2}} \approx 0.4294398.
\end{equation}
Indeed, the value $\min\limits_{x}[x/f(x)]$ is attained when
$f(x) - xf'(x) = 0$, which is equivalent to
\begin{eqnarray}
%g'_{x}(x,\varepsilon) & = & 0 \rightarrow \nonumber \\
 1 - \varepsilon f'(x) & = &
1 - 10\varepsilon (1-x)^{4}\left[1-(1-x)^{5}\right],
\label{1root}
\end{eqnarray}
after replacing $\varepsilon$ by $x/f(x)$.

Formula \eqref{1root2} (and its natural generalization) is missing in
\cite{RU1,SpaCoup2010}, although it simplifies analysis of
$\varepsilon_{0}$.

The minimizating value $x_{0} \approx 0.26057$  in
\eqref{1root2} is the unique root of the equation
\begin{equation}\label{1root1}
(1-x)^{5} + 10x(1-x)^{4} - 1 =0,
\end{equation}
and $\varepsilon_0$ is the single-system threshold.

We now consider the coupled case, and use the one-dimensional potential
function from equation (\ref{defpf11}). For regular $(l,r)$-LDPC-codes
the function $U(x)$ from \eqref{defpf11} can be integrated in closed form
and takes the form
\begin{equation}\label{exreg1}
U(x,\varepsilon) = \frac{1}{r}- \frac{(1-x)^{r}}{r}- x(1-x)^{r-1} -
\frac{\varepsilon}{l}\left[1-(1-x)^{r-1}\right]^{l}.
\end{equation}
We want to find the maximal $\varepsilon^{*} = \varepsilon^{*}(l,r)$
such that $U(x,\varepsilon^{*}) \geq 0$, $x \in [0,1]$, which will be the
coupled-system threshold according to \eqref{sst2}. Consider the
case $l \to \infty$ and $l/r \to \alpha$, $\alpha < 1$. We now show that
$\varepsilon^{*}(l,r) = \alpha$. 

First, we set $\varepsilon = \alpha$
and $l = r\alpha$. Then
$$
\begin{gathered}
rU(x,\alpha) = 1\! - \!(1\!-\!x)^{r} \!-\! rx(1\!-\!x)^{r\!-\!1} \!-\!
\left[1\!-\!(1\!-\!x)^{r\!-\!1}\right]^{r\alpha}, \\
U'_{x}(x,\alpha) = (r\!-\!1)(1\!-\!x)^{r\!-\!2}\left\{x\! -\!
\alpha\left[1\!-\!(1\!-\!x)^{r\!-\!1}\right]^{r\alpha\!-\!1}\right\}.
\end{gathered}
$$
Since $U(1,\alpha) = 0$ and $U'_{x}(x,\alpha) > 0$, $x \geq \alpha$,
we need to consider only $x < \alpha$.

Small values of $x$ can also be exclude as follows. We have
$$
\begin{gathered}
x - \alpha\left[1-(1-x)^{r-1}\right]^{r\alpha-1} \geq
x - \alpha[(r-1)x]^{r\alpha-1} \geq 0, \\
x \leq x_{0} = \frac{1}{(r-1)}
\left[\frac{1}{\alpha(r-1)}\right]^{1/(r\alpha-2)},
\end{gathered}
$$
where $x_{0} \geq 1/(2r)$, and $U'_{x}(x,\alpha) \geq 0$, if
$x \leq 1/(2r)$. Since $U(0,\alpha) = 0$, the interval that remains to
be considered is $x = b/r$, $1/2 < b < \alpha r$. Since $(1-x)^{r} \leq e^{-rx}$
and $(1-x)^{r} \geq e^{-rx/(1-x)}$, we may use the following bounds in
the interval $1/2 < b < \alpha r$
$$
\begin{gathered}
rU(b/r,\alpha) \geq 1- e^{-b} - be^{-b} -
\left[1-e^{-b/(1-\alpha)}\right]^{r\alpha} \geq o(1)
\end{gathered}
$$
for $r \to \infty$. But now
$$
\begin{gathered}
\inf_{0 \leq x \leq 1}U(x,\alpha) = o(1/r), \qquad r\to \infty,
\end{gathered}
$$
and we obtain
\medskip

\begin{proposition}
\label{prop:LDPC-BEC}
For the ensemble of regular $(l,r)$-LDPC codes of rate
$1-l/r$ with $l/r \to \alpha$ as $r \to \infty$, where $\alpha < 1$, the
coupled-system threshold
$\lim\limits_{r \to \infty}\varepsilon^{*}(l,r) = \alpha$.
\end{proposition}
\medskip

Proposition \ref{prop:LDPC-BEC} immediately reveals the important 
\begin{corollary}
\label{cor:LDPC-BEC}
The ensemble of coupled regular $(l,r)$-LDPC codes with
$l/r \to \alpha$ as $r \to \infty$ achieves the capacity 
$1-\alpha$ of the
binary erasure channel with erasure rate $\alpha$.
\end{corollary}
\medskip

Note: In Corollary \ref{cor:LDPC-BEC} we have rederived an
important result from \cite{SpaCoup2010} by elementary methods
from Theorem \ref{th:two} {\em without} the need
for the concept of {\em threshold saturation} or the use of the
area theorem.

\subsection{Multiuser Cancelation}

In \cite{SchTru2012} an iterative interference cancelation system is
discussed with the following 1-dimensional dynamical system equation
\begin{equation}\label{scheme2}
x^{(l+1)} = \alpha g(x^{(l)}) + \sigma^{2}, \qquad l = 0,1,2,\ldots
\end{equation}
where $g(x) > 0$ is a given bounded function and $\sigma \geq 0$ is
a constant, the root of the normalized noise variance. We are interested in the
maximum $\alpha_{0} = \alpha_{0}(g,\sigma)$, such that
$x^{(l)} \to x^{(\infty)} = x^{(\infty)}(g,\sigma)$ as $l \to \infty$.
It is straightforward to show that $x^{(0)},x^{(1)},x^{(2)}, \ldots$
is a monotonically decreasing sequence, i.e.,
\begin{equation}\label{scheme2a}
x^{(l+1)} = \alpha g(x^{(l)}) + \sigma^{2} \leq x^{(l)},
\qquad l = 0,1,2,\ldots
\end{equation}

In order to find stable points of the system \eqref{scheme2}
consider the equation
\begin{equation}\label{scheme2c}
\alpha g(x) + \sigma^{2} - x = 0.
\end{equation}

The values $x_{0} = x^{(\infty)}$ and $\alpha_{0})$ defining a stable point
satisfy the equations
\begin{equation}\label{scheme2d}
\begin{gathered}
\alpha_{0} g(x_{0}) + \sigma^{2} - x_{0} = 0, \\
\alpha g'(x_{0}) - 1 = 0.
\end{gathered}
\end{equation}
Therefore
\begin{equation}\label{scheme2e}
\alpha_{0} = \frac{x_{0} - \sigma^{2}}{g(x_{0})} \geq
\min_{x \in S} \frac{x - \sigma^{2}}{g(x)},
\end{equation}
where $S$ is the set of stationary points of the function
$(x-\sigma^{2})/g(x)$, $x > \sigma\sigma^{2}$, i.e. roots of the
equation
\begin{equation}\label{scheme2f}
g(x) - (x^{2} - \sigma^{2})g'(x) = 0.
\end{equation}

In \cite{TruSch2013} spatial coupling is applied to this system, and,
using the Lyapunov function \eqref{basic42k} with \eqref{basic42k1},
it is shown that the coupled system can approach the
capacity of the multiple access channel.

\section{Approximations}

In \cite{TTK1} certain approximations for behavior of the system
\eqref{2coup1} via partial differential equations were proposed.
We present different approximations here.

Consider the system \eqref{22coup1}, i.e. the equation
\begin{equation}\label{Pf3}
x_{i}^{(l+1)} = f\left(g\left(y_{i}^{(l)}\right)\right), \qquad
i \in {\mathcal L}_{0} = \{-L,\ldots,L\},
\end{equation}
where
$$
y_{i}^{(l)} = \frac{1}{w^{2}}\sum\limits_{k=0}^{w-1}
\sum\limits_{j=0}^{w-1}x_{i+j-k}^{(l)}.
$$

Consider first the case $w-1-L \leq i \leq 1-w + L$.
Note that
$$
y_{i}^{(l)} = \frac{1}{w^{2}}
\sum\limits_{m=-(w-1)}^{w-1}a(m)u_{i+m}^{(l)},
$$
where $a(m)$ is the number of solutions of the equation
$j-k=m$, $0 \leq j,k \leq w-1$, i.e. $a(m) = w-|m|$. Then
\begin{equation}\label{Pf31}
y_{i}^{(l)} \approx \frac{1}{w^{2}}\int\limits_{-w}^{w}
(w-|r|)x_{i+t}^{(l)}dr.
\end{equation}
On the accuracy of the approximation \eqref{Pf31}
the following inequality holds
\begin{equation}\label{Pf31f}
\left|y_{i}^{(l)} - \frac{1}{w^{2}}\int\limits_{-w}^{w}
(w-|r|)x_{i+t}^{(l)}dr\right| \leq \frac{1}{w}.
\end{equation}
Denoting
$$
v_{i}^{(l+1)} = g^{-1}f^{-1}\left(x_{i}^{(l+1)}\right), \qquad
x_{i}^{(l)} = f\left(g\left(v_{i}^{(l)}\right)\right),
$$
we obtain from \eqref{Pf3} the non-linear integral equation
$$
v_{i}^{(l+1)} \approx \frac{1}{w^{2}}\int\limits_{-w}^{w}
(w-|r|)f\left(g\left(v_{i+r}^{(l)}\right)\right)dr.
$$
Changing variables $r = sw$, $i=wx$ and denote $\alpha = L/w$,
we obtain for $1-1/w-\alpha \leq x \leq \alpha +1/w-1$
$$
v_{x}^{(l+1)} \approx \int\limits_{-1}^{1}
(1-|s|)f\left(g\left(v_{x+s}^{(l)}\right)\right)ds.
$$
Denoting $v_{x}^{(t)} = v(x,t)$, we get the approximation
\begin{equation}\label{Pf31a}
\frac{\partial v(x,t)}{\partial t} \approx \int\limits_{-1}^{1}
(1-|s|)f\left(g\left(v(x+s,t)\right)\right)ds - v(x,t).
\end{equation}
For the fixed points $v(x)$ of this equation we obtain
($1-1/w-\alpha \leq x \leq \alpha +1/w-1$) and
\begin{equation}\label{Pf31b}
\int\limits_{-1}^{1}(1-|s|)f\left(g\left(v(x+s)\right)\right)ds =
v(x).
\end{equation}

Consider the case $-L \leq i \leq w-1-L$ (i.e. the left boundary).
Then, analogously, we obtain
$$
\begin{gathered}
y_{i} \approx \frac{1}{w^{2}}\int\limits_{-(L+i)}^{w}
(w-|r|)x_{i+r}^{(l)}dr
\end{gathered}
$$
and ($-\alpha \leq x \leq -\alpha-1/w+1$)
\begin{equation}\label{Pf31c}
\begin{gathered}
\frac{\partial v(x,t)}{\partial t} \approx \\
\int\limits_{-(\alpha +x)}^{1}
(1-|s|)f\left(g\left(v(x+s,t)\right)\right)ds - v(x,t).
\end{gathered}
\end{equation}
For the fixed points $v(x)$ of that equation we get
\begin{equation}\label{Pf31d}
\int\limits_{-(\alpha +x)}^{1}(1-|s|)f\left(g\left(v(x+s)\right)
\right)ds = v(x).
\end{equation}

Similar approximations for the right boundary can be obtained.

\section{Conclusion}

\bibliographystyle{IEEEtran}

\end{document}